\documentclass[twocolumn,showpacs,superscriptaddress,prb,floatfix]{revtex4}
\usepackage{graphicx, amsfonts, array, gensymb}
\begin{document}
\author{O. Leenaerts}
\email{ortwin.leenaerts@ua.ac.be} \affiliation{Universiteit Antwerpen, Departement Fysica,
Groenenborgerlaan 171, B-2020 Antwerpen, Belgium}
\author{H. Peelaers}
\email{hartwin.peelaers@ua.ac.be} \affiliation{Universiteit Antwerpen, Departement Fysica,
Groenenborgerlaan 171, B-2020 Antwerpen, Belgium}
\author{A. D. Hern\'andez-Nieves}
\email{alexande@cab.cnea.gov.ar} \affiliation{Universiteit Antwerpen, Departement Fysica,
Groenenborgerlaan 171, B-2020 Antwerpen, Belgium}
\affiliation{Centro Atomico Bariloche, 8400 San Carlos de Bariloche,
Rio Negro, Argentina}
\author{B. Partoens}
\email{bart.partoens@ua.ac.be} \affiliation{Universiteit Antwerpen, Departement Fysica,
Groenenborgerlaan 171, B-2020 Antwerpen, Belgium}
\author{F. M. Peeters}
\email{francois.peeters@ua.ac.be} \affiliation{Universiteit Antwerpen, Departement Fysica,
Groenenborgerlaan 171, B-2020 Antwerpen, Belgium}
\date{\today}
\title{First-principles investigation of graphene fluoride and graphane}

\begin{abstract}
Different stoichiometric configurations of graphane and graphene fluoride are investigated within density functional theory. Their structural and electronic properties  are compared, and we indicate the similarities and differences among the various configurations. Large differences between graphane and graphene fluoride are found that are caused by the presence of charges on the fluorine atoms. A new configuration that is more stable than the boat configuration is predicted for graphene fluoride. We also perform GW calculations for the electronic band gap of both graphene derivatives. These band gaps and also the calculated Young's moduli are at variance with available experimental data. This might indicate that the experimental samples contain a large number of defects or are only partially covered with H or F.
\end{abstract}

\pacs{61.48.Gh ,68.43.-h, 73.22.Pr, 81.05.ue} \maketitle

\section{Introduction}

Two-dimensional crystals have been given a large amount of attention since the isolation of one-atom-thick 
materials by Novosolov \textit{et al}.\ in 2004.\cite{nov04,nov05} Graphene, a single layer of graphite, has attracted by far the most attention because of the high crystal quality of the graphene samples and its fascinating electronic properties.\cite{gei07} These properties make it a 
promising candidate to use as a basic material for future electronics applications.\cite{tro07} 
However, the use of graphene for applications in electronics suffers from a major drawback: graphene is, in its pristine state, a 
zero-bandgap semiconductor and this gapless state appears to be rather robust. Several ways have been explored to induce a 
finite band gap in graphene. It was found experimentally that a band gap can be opened by confining the electrons in 
nanoribbons \cite{han07} or by applying a potential difference over a graphene bilayer.\cite{oht06,mcc06} 

The chemical modification of graphene is another promising way to create a band gap. \cite{eli09,che10, wit10, nai10}
When radicals such as oxygen, hydrogen, or fluorine atoms are adsorbed on the graphene surface they form covalent bonds with the 
carbon atoms. These carbon atoms change their 
hybridization from $sp^2$ to $sp^3$, which leads to the opening of a band gap (similar as in diamond). The adsorbed radicals can 
attach to the graphene layer in a random way, as is the case in graphene oxide (GO),\cite{dik07, eda10} or they can form ordered 
patterns. In the last case,  new graphene-based 2D crystals are 
formed with properties that can vary greatly from their parent material. This has been found to be the case for hydrogen 
and fluorine adsorbates. The new 2D crystals that are expected to form in those cases\cite{slu03} have been named graphane \cite{eli09,sof07} 
and graphene fluoride (or fluorographene)\cite{nai10}, respectively. 

Following this route, multi-layer graphene fluoride was recently synthesized, \cite{che10, wit10}  and its structural and electronic properties were studied.
A strongly insulating behavior was found with a room temperature resistance larger than 10G$\Omega$, which is consistent with the 
existence of a large band gap in this new material.\cite{che10, wit10} Only a partial fluorine coverage of the graphene multi-layer samples was achieved
in these experiments. The F/C ratio was estimated to be 0.7 in Ref. \onlinecite{che10} and 0.24 in 
Ref. \onlinecite{wit10}, according to weight gain measurements.

An important step forward in creating fully covered two-dimensional graphene fluoride samples was recently achieved in 
Ref. \onlinecite{nai10}. The obtained single-layer graphene fluoride exhibits a strong insulating behavior with a room 
temperature resistance larger than 1T$\Omega$, a strong temperature stability up to 400 $\celsius$, and almost a complete 
disappearance of the graphene Raman peaks associated with regions that are not fully fluorinated.\cite{nai10} 
The graphene Raman peaks do not disappear completely, however, which could be an indication of the presence of defects in the sample, 
such as a small portion of carbon atoms not bonded to fluorine atoms. It was also found experimentally that fluorographene has 
a Young's modulus of $\approx$ 100 N/m, and the optical measurements suggest a band gap of $\approx$ 3eV. 

In Ref.\ \onlinecite{rob10} it was demonstrated that single-side adsorption is also possible and that it probably results in a crystalline C$_4$F structure with a large band gap.

On the theoretical side, first-principles studies on graphene monofluoride started in 1993, motivated by available experiments on graphite monofluoride. 
Using density functional theory (DFT) calculations, it was shown in Ref. \onlinecite{cha93} that the chair configuration of graphene fluoride is energetically more favorable than 
the boat configuration by 0.145 eV per CF unit (0.073 eV/atom), while a transition barrier of the order of 2.72 eV was found 
between both structures. Due to the small difference in formation energy and the large energy barrier between both configurations, it was 
argued that the kinematics of the intercalation could selectively determine the configuration, or that there could also be a mixing of 
both configurations in the available experiments. By using the local density approximation (LDA) for the 
exchange-correlation functional a direct band gap of 3.5 eV was calculated for the chair configuration in 
Ref. \onlinecite{cha93}. 
However, it is well known that DFT generally underestimates the band gap. Recent calculations 
used the more accurate GW approximation and found a much larger band gap of 7.4 eV for the chair configuration of 
graphene monofluoride (Ref. \onlinecite{kli10}). This theoretical value is twice as large as the one obtained experimentally
for graphene fluoride in Ref. \onlinecite{nai10}, which is $\approx$ 3eV. The experimental value for the Young's modulus as found in
Ref. \onlinecite{nai10} ($\approx$ 100 N/m) is also half the value obtained recently from first-principles 
calculations in Ref. \onlinecite{mun10} ($\approx$ 228 N/m) for the chair configuration of graphene fluoride. It is worth noting 
that the experimental \cite{lee08} and theoretical \cite{liu07} values of the Young's modulus of graphene only differ in a small 
percentage. 

Possible reasons for the disagreement between the experimental values and the ab initio results for the Young's modulus and the 
band gap of graphene fluorine could be: i) the presence of a different configuration or a mixture of them in the experimental samples, 
or ii) the presence of defects, which could decrease the size of both the Young's modulus and the band gap from the expected theoretical values.

In this paper, we investigate various possible crystal configurations for both graphene-based two-dimensional crystals, 
graphene fluoride and graphane, and we examine their structural, electronic, and mechanical properties. In the case of graphene 
fluorine, we found a new configuration not considered before that has a lower energy than the boat configuration. This new 
configuration, which we call the zigzag configuration, is energetically less favorable than the chair 
configuration by only 0.073 eV per CF unit (0.036 eV/atom). We calculated the Young's modulus and the band gap (both with GGA and 
in the GW approximation) for the different configurations. The disagreements between experimental and ab initio calculations for graphene 
fluoride persist independently of the considered configuration. These results imply that the available experimental samples probably contain a large 
number of defects, such as a portion of carbon atoms not bonded to fluorine atoms, that decrease the value of both the Young's 
modulus and the band gap from the expected theoretical values.

The paper is organized as follows: first we describe the computational details of our first-principles calculations. Then we investigate the stability and structural properties of the different configurations of both graphene derivatives. To conclude, the elastic and electronic properties of the different structures are discussed.  

\section{Computational details}\label{comp_det}
We examine different graphane and graphene fluoride configurations with the use of \textit{ab initio} calculations performed within the density functional theory (DFT) formalism. The generalized gradient approximation (GGA) of Perdew, Burke, and Ernzerhof (PBE)\cite{per96} is used for the exchange-correlation functional and  a plane wave basis set with a cutoff energy of 40 Hartree is applied.  The sampling of the Brillouin zone is done with the equivalent of a $24\times 24 \times 1$ Monkhorst-Pack k-point grid\cite{mon71} for a graphene unit cell and we use pseudopotentials of the Troullier-Martins type.\cite{tro91} Since periodic boundary conditions are applied in all 3 dimensions a vacuum layer of 20 Bohr is included to minimize the (artificial) interaction between adjacent layers. All the calculations were performed with the \textsc{abinit} code.\cite{abi02} 

The reported quasiparticle corrections for the band gap are obtained using the \textsc{yambo} code.\cite{yam09} Here the first-order quasiparticle corrections are obtained using Hedin's GW approximation~\cite{hed65} for the electron self-energy. Because we are treating two-dimensional systems, the spurious Coulomb interaction between a layer and its images should be avoided, as this causes serious convergence problems. Therefore we use a truncation of this interaction in a box layout, following the method of Rozzi \emph{et al.}.\cite{roz06} The remaining singularity is treated using a random integration method in the region near the gamma
point.\cite{yam09} Nevertheless, a larger separation between the layers is necessary, so a value of $60$ Bohr is used for these calculations.

\section{Results}
\begin{figure}[h!]
  \centering
\includegraphics[width= 3.375 in]{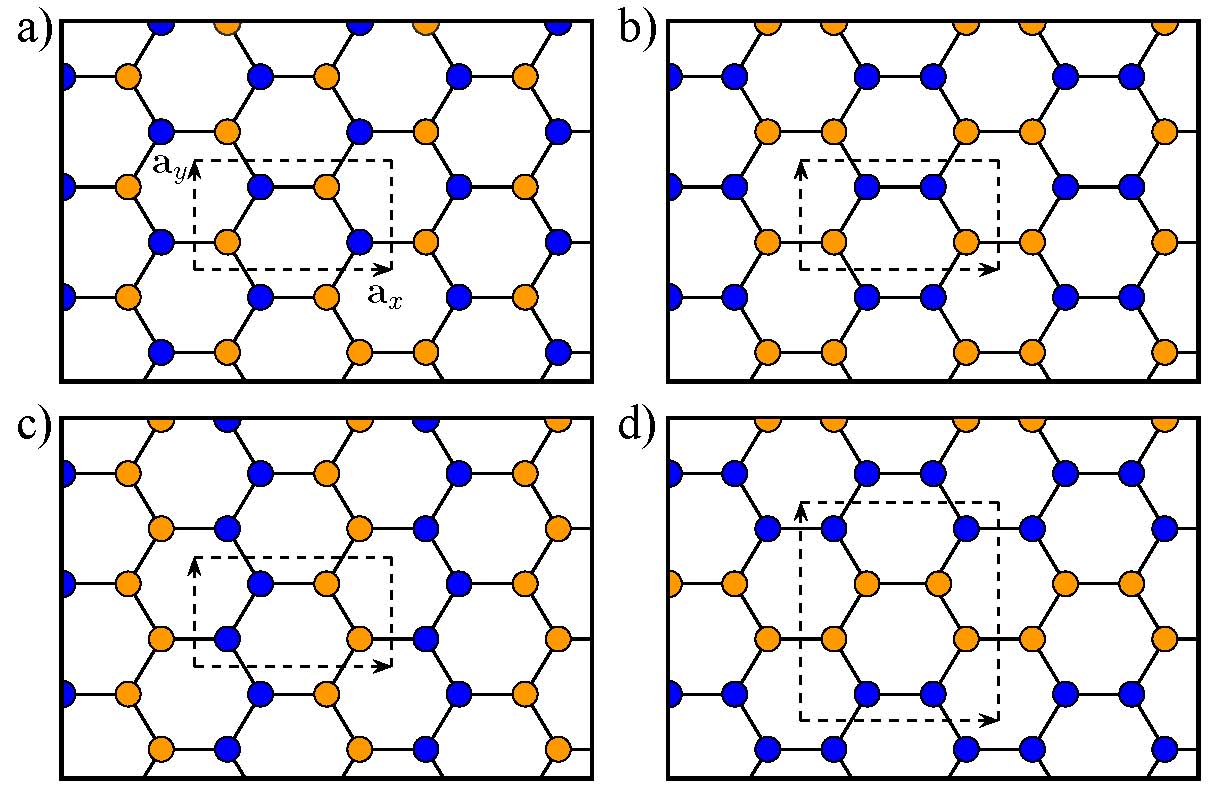}
\caption{(Color online) Four different configurations of hydrogen/fluorine-graphene: a) chair, b) boat, c) zigzag, and d) armchair configuration. The different colors (shades) represent adsorbates (H or F) above and below the graphene plane. The supercell used to calculate the elastic constants is indicated by the dashed box.\label{fig_conform}}
\end{figure}

We studied four different stoichiometric configurations for both graphane and graphene fluoride in which every carbon atom is covalently bonded to an adsorbate in an equivalent way, i.e.\ every carbon/adsorbate pair has the same environment. These configurations are schematically depicted in Fig.\ \ref{fig_conform} and we will refer to them as the `chair', `boat', `zigzag', and `armchair' configuration. The chair and boat configurations have been well investigated before, but the zigzag and armchair configurations are rarely examined for graphane\cite{slu03} and we are not aware of any studies for fluorographene. The names of these last two configurations have been chosen for obvious reasons (see Fig. \ref{fig_conform}(c) and \ref{fig_conform}(d)). After relaxation, the different configurations appear greatly distorted when compared with the schematic pictures of  Fig.\ \ref{fig_conform}, so these figures should only be regarded as topologically correct (see Fig.\ \ref{fig_superlatt}). 

\subsection{Stability analysis}
To examine the stability of the different configurations, we make use of the formation energy of the structures and the binding energy between the graphene layer and the adsorbates. We define the formation energy, E$_{\text{f}}$, as the energy per atom of the hydrogenated or fluorinated graphene with respect to intrinsic graphene and the corresponding diatomic molecules H$_2$ and F$_2$. The binding energy, E$_{\text{b}}$, is defined  with respect to graphene and the atomic energies of the adsorbates and is calculated per CH or CF pair. The results are summarized in Table \ref{tab_energies}.

\begin{table}[h] \centering
\caption{The formation energy E$_f$, the binding energy E$_b$, and the relative binding energy $\Delta$E$_f$ (with respect to the most stable configuration) for different hydrogenated and fluorinated graphene configurations. The energies are given in eV.\label{tab_energies}}
\begin{tabular}{ccccc}
\hline\hline
 & chair & boat & zigzag & armchair\\
\hline
  graphane								&   &   &    \\
	E$_{\text{b}}$					& -2.481 & -2.378 & -2.428 & -2.353   \\
  E$_{\text{f}}$					& -0.097 & -0.046 & -0.071 & -0.033  	\\
  $\Delta$E$_{\text{f}}$	&  0.000 &  0.051 &  0.027 &  0.064  	\\
  fluorographene					&   &   &    \\
	E$_{\text{b}}$					& -2.864 & -2.715 & -2.791 & -2.673   \\
  E$_{\text{f}}$					& -0.808 & -0.733 & -0.772 & -0.712   \\
  $\Delta$E$_{\text{f}}$	&  0.000 &  0.075	&  0.036 &  0.095   \\
\hline\hline
\end{tabular}
\end{table}

As has been reported before, the chair configuration is the most stable one for both graphane\cite{sof07} and graphene fluoride\cite{han10}. The zigzag configuration is found to be more stable than the boat and armchair configurations and its formation energy is only slightly higher than that of the chair configuration: for both graphene derivatives the difference in formation energy, $\Delta$E$_{\text{f}}$, between chair and zigzag is of the order of the thermal enenrgy at room temperature (26 meV). The energy differences between the various configurations are more pronounced for graphene fluorine than for graphane but they are of the same order of magnitude. 

When we compare graphane and fluorographene, the binding energy of hydrogen and fluorine appears to be rather similar (2.5 eV compared to 2.9 eV) but there is a huge difference in the formation energy (0.1 eV compared to 0.9 eV). This is a consequence of the large difference in the dissociation energy between hydrogen and fluorine molecules. The formation energy as defined above can be regarded as a measure of the stability against molecular desorption from the graphene surface. Therefore graphene fluoride is expected to be much more stable than graphane as has indeed been observed experimentally.\cite{eli09,nai10}\\

\subsection{Structural properties}

Besides the large difference in formation energy there are also pronounced structural differences between both graphene derivatives. The structural parameters for the different configurations of graphane and fluorographene are shown in Table \ref{tab_struct}. Note that all the structures are described  in an orthogonal supercell, as illustrated in Fig.\ \ref{fig_conform}, for ease of comparison. The results for the chair configuration agree well with previous theoretical calculations for graphane \cite{sof07,flo09,me09} and graphene fluoride.\cite{han10}

\begin{table}[h] \centering
\caption{Structure parameters for the different hydrogenated and fluorinated graphene derivatives. Distances are given in $\text{\AA}$ and angles in degrees.\label{tab_struct}
 The distance between neighboring C atoms, $d_{\textsc{cc}}$, and the angles, $\theta_{\textsc{ccx}}$, are averaged over the supercell.}
\begin{tabular}{ccccc}
\hline\hline
 & chair & boat & zigzag & armchair\\
\hline
  graphane														&   		&   		&   		& 				\\

  $a_x/\sqrt{3}$											& 2.539 & 2.480 & 2.203 & 2.483   \\
  $a_y/n_y$														& 2.539 & 2.520 & 2.540 & 2.270   \\
	$d_{\textsc{ch}}$  	 								& 1.104 & 1.099 & 1.099 & 1.096   \\
  $\overline{d}_{\textsc{cc}}$				& 1.536 & 1.543 & 1.539 & 1.546  	\\
  $\overline{\theta}_{\textsc{cch}}$	& 107.4 & 107.0	& 106.8	& 106.7		\\
  $\overline{\theta}_{\textsc{ccc}}$ 	& 111.5	& 111.8	& 112.0	& 112.1		\\

  fluorographene											&   		&   		&   		& 				\\

  $a_x/\sqrt{3}$											& 2.600 & 2.657 & 2.415 & 2.662   \\
  $a_y/n_y$														& 2.600 & 2.574 & 2.625 & 2.443   \\
	$d_{\textsc{cf}}$  	 								& 1.371 & 1.365 & 1.371 & 1.365   \\
  $\overline{d}_{\textsc{cc}}$				& 1.579 & 1.600 & 1.585 & 1.605  	\\
  $\overline{\theta}_{\textsc{ccf}}$	& 108.1 & 106.0	& 104.6	& 104.2		\\
  $\overline{\theta}_{\textsc{ccc}}$ 	& 110.8	& 112.8	& 113.9	& 114.2		\\
    \hline\hline
\end{tabular}
\end{table}

It is also useful to compare the interatomic distances and bond angles with those of graphene and diamond. Therefore we calculated these using the same formalism as described above (section \ref{comp_det}). The C-C bond has a length of 1.42 $\text{\AA}$ for graphene compared to 1.54 $\text{\AA}$  for diamond, and the bond angles are  120$^{\circ}$ and 109.5$^{\circ}$, respectively. Notice that both graphane and fluorographene resemble much closer the diamond structure than graphene, which is not surprising since the hybridization of the carbon atoms in these structures is the same as in diamond, i.e.\ sp$^3$. The C-C bond length for the graphane configurations is similar to the one in diamond, but $\overline{d}_{\textsc{cc}}$	in fluorographene is about 0.05 $\text{\AA}$ larger. This can be explained from a chemical point of view as due to a depopulation of the bonding orbitals between the carbon atoms.  The depopulation of these bonding orbitals results from an electron transfer from the carbon to the fluorine atoms due to the difference in electronegativity between C and F. We used a Hirshfeld-based method\cite{me08,bul07,hir77} to calculate this charge transfer and found it to be $\Delta Q\approx0.3e$. The charge transfer in graphane is much smaller because of the similarity between the electron affinity of C and H.

\begin{figure}[h!]
  \centering
\includegraphics[width= 3.375 in]{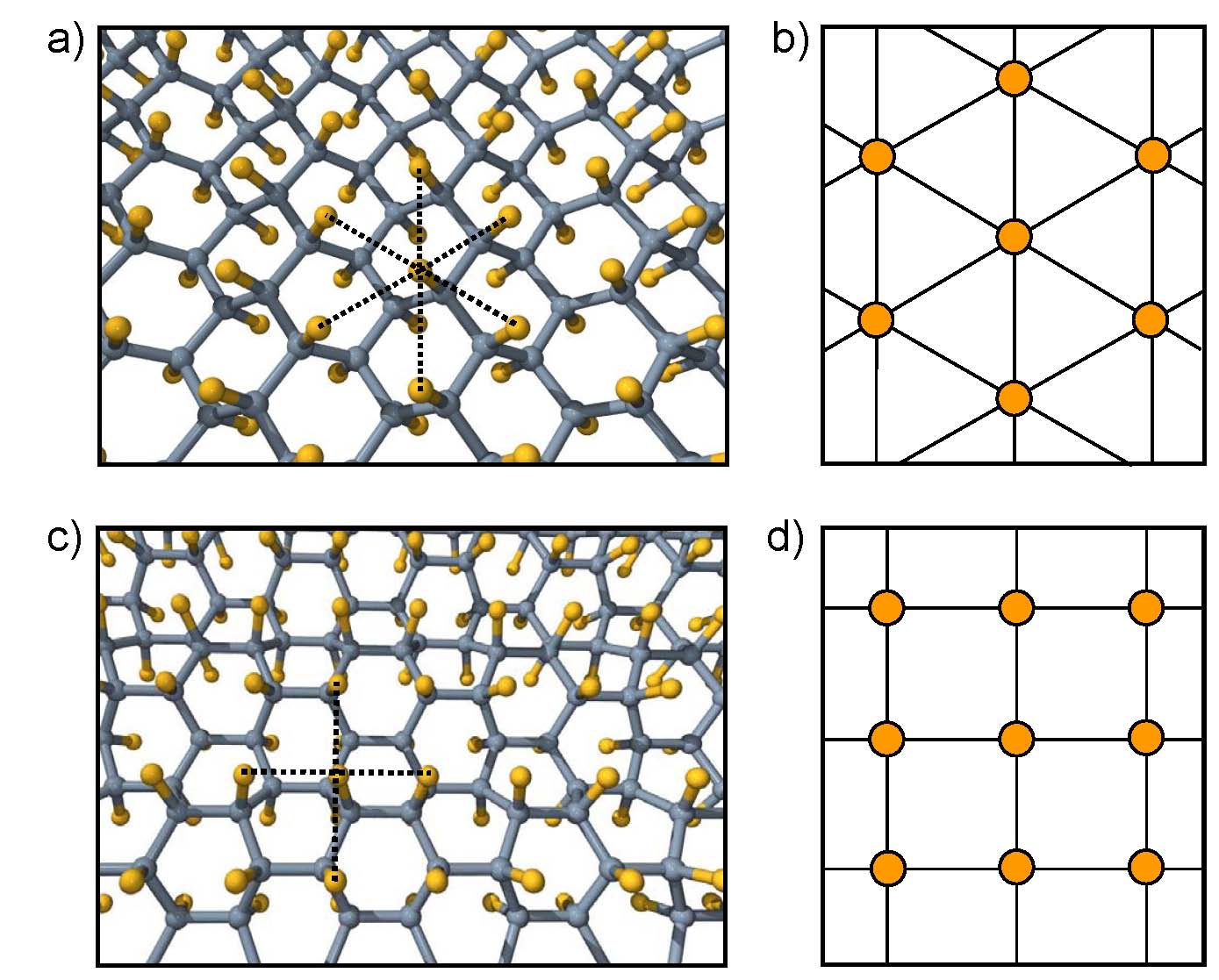}
\caption{(Color online) Fluorographene in the zigzag (a) and armchair (c) configuration. The nearest neighbor bonds of one F atom are indicated with dotted lines to show the symmetry of the superlattices (as shown in (b) and (d)). b) The hexagonal superlattice which is formed in case of chair and zigzag configurations. d) The cubic superlattice which is formed in case of boat and armchair configurations.\label{fig_superlatt}}
\end{figure}

The fact that the fluorine atoms are negatively charged has an appreciable influence on the structure of graphene fluoride when compared to graphane. This can, e.g., be seen from the sizes of the different bond angles. The bond angles (and also the bond lengths) in the chair configuration can be regarded as the ideal angles (lengths) for these structures because they can fully relax. The other configurations will try to adopt these ideal bond angles and it can be seen from Table \ref{tab_struct} that this is indeed the case for the graphane configurations. The fluorographene configurations, on the other hand, appear to be somewhat distorted because their bond angles are (relatively) far from ideal. This is probably caused by the repulsion between the different fluorine atoms as can be demonstrated when focusing only on the positions of the F atoms. The fluorine atoms appear to form hexagonal or cubic superlattices depending on the configuration (see Figs.\ \ref{fig_superlatt}(b) and \ref{fig_superlatt}(d)). This is trivial in the case of the chair (and maybe the boat) configuration  but not so for the others (see Figs.\ \ref{fig_superlatt}(a) and \ref{fig_superlatt}(c)). These superlattices are not perfect (deviations of a few percent), but are much more pronounced than in the case of graphane. 

So it seems that, at the cost of deforming the bonding angles, F superlattices are formed to minimize the electrostatic repulsion between the charged F atoms.

\subsection{Elastic strain}
Graphene and its derivatives graphane and fluorographene can be isolated and made into free-hanging membranes. This makes it possible to measure the elastic constants of these materials from nanoindentation experiments using an atomic force microscope.\cite{lee08,eli09,nai10} The experimental elastic constants can be compared to first-principles calculations which gives us information about the purity and structural crystallinity of the experimental samples. Therefore we calculated the (2D) Young's modulus, $E'$, and the Poisson's ratio, $\nu$, of the different graphane and fluorographene configurations, which we list in Table \ref{tab_elast}. The Young's modulus and the Poisson's ratio of graphene are found to be $E'=336$ Nm$^{-1}$ and $\nu=0.17$, respectively, which corresponds well to the experimental value, $E'_{\text{exp}}=340\pm50$ Nm$^{-1}$, and other theoretical results.\cite{top10,mun10} 

\begin{table}[h] \centering
\caption{\label{tab_elast} Elastic constants of the different hydrogenated and fluorinated graphene derivatives. The 2D Young's modulus, $E'$, and Poisson's ratio, $\nu$, are given along the cartesian axes. $E'$ is expressed in Nm$^{-1}$.}
\begin{tabular}{ccccc}
\hline\hline
 & chair & boat & zigzag & armchair\\
\hline
  graphane			&   &   &  &  \\
  $E'_x$ 				&  243  &  230  &  117 &  247   \\
	$E'_y$ 				&  243  &  262  &  271 &  142   \\
  $\nu_x$ 		 	& 0.07  & -0.01 & 0.05 & -0.05 	\\
  $\nu_y$ 		 	& 0.07  & -0.01 & 0.11 & -0.03 	\\  
  fluorographene&   &   &  &  \\
  $E'_x$   			&  226  &  238  &  240 &  215  	\\
	$E'_y$   			&  226  &  240  &  222 &  253  	\\
  $\nu_x$ 		 	&  0.10 &  0.00 & 0.09 &  0.02  \\
  $\nu_y$ 		 	&  0.10 &  0.00 & 0.11 &  0.02  \\
\hline\hline
\end{tabular}
\end{table}

The 2D Young's modulus of graphane and fluorographene is smaller than that of graphene. The $E'$ of the chair and boat configurations of both graphene derivatives are about 2/3 the value of graphene which makes them very strong materials. The Young's modulus for the zigzag and armchair configurations of graphane are highly anisotropic with values that are roughly halved along the direction that shows the largest crumpling (see Figs.\ \ref{fig_superlatt} (a) and \ref{fig_superlatt}(c)). The situation is completely different for fluorographene where the Young's modulus is more isotropic. The values that are found for the chair configurations agree well with recent calculations (245 Nm$^{-1}$ and 228 Nm$^{-1}$ for graphane and fluorographene, respectively, in Ref.\ \onlinecite{mun10}). Nair \textit{et al.} performed a nanoindentation experiment on fluorographene\cite{nai10} and measured a value of $100\pm30$ Nm$^{-1}$ for $E'_{\textsc{fg}}$. This value is approximately half the theoretical value, which suggests that the experimental samples contain a lot of defects.\cite{top10}

The Poisson's ratio shows a similar behavior as the Young's modulus. The knowledge of $E'$ and $\nu$ allows us to calculate all the other 2D elastic constants\cite{tho85} such as the bulk, $K'=E'/2(1-\nu)$, and shear modulus, $G'=E'/2(1+\nu)$. For the chair configurations we find $K'_{\text{HG}}=131$ Nm$^{-1}$ and $G'_{\text{HG}}=114$ Nm$^{-1}$ for graphane, and $K'_{\text{FG}}=126$ Nm$^{-1}$ and $G'_{\text{FG}}=103$ Nm$^{-1}$ for graphene fluoride.

\subsection{Electronic properties}
Graphene is a zero-gap semiconductor but its derivatives, like graphane and fluorographene, have large band gaps, similar to diamond. In Table \ref{tab_bg}, the band gaps of the configurations under study are given. We also performed GW calculations because GGA is known to underestimate the band gap. The GGA and GW results show different behavior for the variation of the band gap among the different configurations. Note that this indicates that it is not straightforward to deduce qualitative trends from GGA as is often done in the literature. But, overall, we may conclude that the band gap is more or less independent of the configuration and that its size is roughly twice as large for GW as compared to GGA.

The GGA results give a band gap of 3.2 eV for the most stable fluorographene configuration which is in accordance with the experimental result of $\sim$3 eV as found in Ref.\ \onlinecite{nai10}. However, this value is much smaller than the (more accurate) GW band gap of 7.4 eV, so that the theoretical and experimental results differ by about a factor of two. This conflict might be resolved if the experimental value is ascribed to midgap states due to defects in the system, such as missing H/F atoms (similar to what has been predicted for defected graphane\cite{bre10}). 

\begin{table}[h] \centering
\caption{Electronic properties of the different configurations of hydrogenated and fluorinated graphene derivatives. The electronic band gap, $E_{\text{gap}}$, is given for GGA and GW calculations. The ionization energy (IE) is also calculated. All the energies are given in eV.\label{tab_bg} }
\begin{tabular}{ccccc}
\hline\hline
 & chair & boat & zigzag & armchair\\
\hline
  graphane			& & & & \\
  E$_{\text{gap}}$ (GGA)				    &  3.70  &  3.61  &  3.58  &  3.61  \\
	E$_{\text{gap}}$ (GW)	 		    		&  6.05  &  5.71  &  5.75  &  5.78  \\
  IE (GGA)													&  4.73  &  4.58  &  5.30  &  4.65  \\
	
  fluorographene& & & & \\
  E$_{\text{gap}}$ (GGA)				    &  3.20  &  3.23  &  3.59  &  4.23  \\
	E$_{\text{gap}}$ (GW)			    		&  7.42  &  7.32  &  7.28  &  7.98  \\
  IE (GGA)													&  7.69  &  7.64  &  7.85  &  8.27  \\
   
\hline\hline
\end{tabular}
\end{table}

\begin{figure}[h!]
  \centering
\includegraphics[width= 3.2 in]{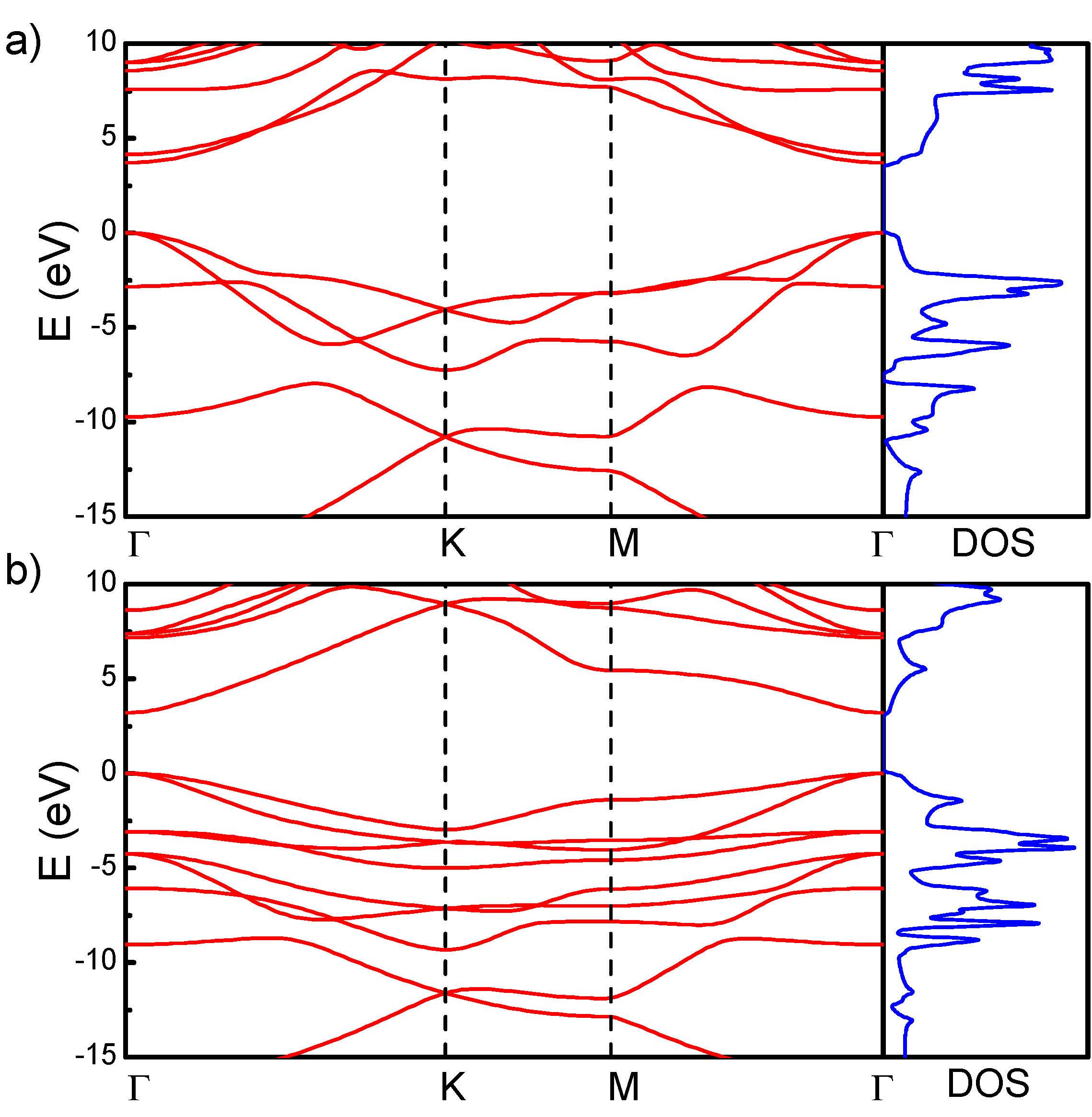}
\caption{(Color online) \label{fig_bs} The electronic band structure and the corresponding density of states (GGA) for the chair configuration of graphane (a) and fluorographene (b). The valence band maximum has been used as the origin of the energy scale.}
\end{figure}

The electronic band structure and the corresponding density of states of graphane and fluorographene in the chair configuration are shown in Fig.\ \ref{fig_bs}. Both band structures look similar but there are also some clear differences. In the case of fluorographene the parabolic band at the $\Gamma$-point, corresponding to quasi-free electron states, is at much higher energies which indicates a larger ionization energy for fluorinated graphene. This ionization energy (IE) is defined as the difference between the vacuum level and the valence band maximum and an explicit calculation of this energy indicates a difference of about 3 eV between graphane and fluorographene (see Table \ref{tab_bg}). This is a consequence of the negative charges on the fluorine atoms in fluorographene.  We can also compare the IE values with the work function of graphene which is the same as its ionization potential (because graphene has no band gap) and has a value of 4.22 eV from GGA (this is somewhat smaller that the experimental value\cite{yu09} of 4.57$\pm$0.05 eV). It can be seen from Table \ref{tab_bg} that the ionization energies of both graphene derivatives are higher than that of graphene ($\approx$ 0.5 eV and 3.5 eV, respectively), although the ionization energy of graphane is rather similar to graphene.

\section{Summary and conclusions}
We investigated different configurations of the graphene derivatives fluorographene and graphane. The chair configuration is the most stable one in both cases, but the zigzag configuration has only a slightly higher formation energy and is more stable than the much more studied boat configuration. Fluorographene is found to be much more stable than graphane which is mainly due to a much higher desorption energy for F$_2$ as compared to H$_2$. We also demonstrated that there are structural and electronic differences that are caused by the charged state of the F atoms in fluorographene. 

When our results are compared to available experimental data for fluorographene some discrepancies can be noticed: for all the configurations studied we find much larger band gaps in the electronic band structure and the calculated Young's modules is much larger. This might indicate that the experimental samples still contain appreciable amounts of defects. The nature of these defects requires further investigation, but one can speculate that these defects consist of missing adsorbates, partial H/F coverage, or mixed configurations.

\begin{acknowledgments}
This work was supported by the Flemish Science Foundation (FWO-Vl), the NOI-BOF of the University of Antwerp,the Belgian Science Policy (IAP) and the collaborative project FWO-MINCyT (FW/08/01). A.D.H. also acknowledges support from ANPCyT (Grant No. PICT 2008-2236) and a FWO-Vl senior postdoc fellowship.
\end{acknowledgments}

\end{document}